\documentclass[11pt,aps,prc,superscriptaddress,showpacs,floatfix]{revtex4-1}
\usepackage[dvips]{graphicx}
\usepackage{tikz-feynhand}
\usepackage{bm}
\usepackage{amssymb}
\usepackage{textcomp}
\usepackage[T1]{fontenc}
\usepackage{wrapfig}
\usepackage{ulem}
\usepackage{multirow}
\usepackage{color}
\usepackage{soul}
\newcommand{\Asla}{\ooalign{\hfil/\hfil\crcr{$A$}}}

\newcommand{\cA}{{\cal A}}
\newcommand{\cE}{{\cal E}}

\newcommand{\cW}{{\cal W}}

\newcommand{\cM}{{\cal M}}

\newcommand{\psibar}{\mbox{$\overline{\psi}$}}

\newcommand{\sgm}{\mbox{$s_{\gamma}$}}

\newcommand{\tldJ}{\mbox{${\tilde J}$}}

\newcommand{\bA}{\bf A}

\newcommand{\vbr}{\mbox{${\bm r}$}}

\newcommand{\vA}{\mbox{\boldmath $A$}}

\newcommand{\valpha}{\mbox{\boldmath $\alpha$}}

\newcommand{\zr}{\mbox{${\bm r}$}}

\begin{document}

\title{Photon vortex generation from nonlinear Compton scattering in Feynman approach}

\author{Tomoyuki~Maruyama}
\email{maruyama.tomoyuki@nihon-u.ac.jp}
\affiliation{College of Bioresource Sciences, Nihon University, Fujisawa 252-0880, Japan }
\affiliation{Tokyo Metropolitan university, Hachioji, Tokyo 181-8588, Japan}
\author{Takehito~Hayakawa}
\email{hayakawa.takehito@qst.go.jp}
\affiliation{Kansai Institute for Photon Science, National Institutes for Quantum Science and Technology,
  Kizugawa, Kyoto 619-0215, Japan}
\affiliation{Institute of Laser Engineering, Osaka University, Suita, Osaka 565-0871, Japan}
\author{Ryoichi~Hajima}
\affiliation{Kansai Institute for Photon Science, National Institutes for Quantum Science and Technology,
  Kizugawa, Kyoto 619-0215, Japan}
\author{Toshitaka~Kajino}
\affiliation{Beihang University, School of Physics, International Research Center for Big-Bang Cosmology and Element Genesis, Peng Huanwu Collaborative Center for Research and Education, Beijing 100191, China}
\affiliation{National Astronomical Observatory of Japan, 2-21-1 Osawa,
  Mitaka, Tokyo 181-8588, Japan}
\affiliation{Graduate School of Science, The University of Tokyo,
7-3-1 Hongo, Bunkyo-ku, Tokyo 113-033, Japan}
\author{Myung-Ki Cheoun}
\affiliation{Department of Physics and OMEG institute,
  Soongsil University, Seoul 156-743, Korea}

\date{\today}

\begin{abstract}
In the present study, we show calculation of nonlinear Compton scattering with circularly polarized photons in a cylindrical coordinate using Feynman diagram to calculate photon vortex generation in intermediate states considering conservation of angular momentum.
We take two different vortex wave functions based on Bessel function for the emitted photon and the electron after emission of the photon, which are the eigenstate of $z$ component of the total angular momentum (zTAM) when a particle propagates along $z$ axis.
The result shows that when an electron absorbs $N$ photons a photon vortex with a zTAM = $N$ is predominantly radiated, but there are still very small contributions of photons with a zTAM = ($N - 1$) and ($N + 1$) due to the spin flip of the initial electron.
This means even when an electron absorbs only a single photon, the electron may emit a photon vortex of a zTAM of 2$\hbar$.
However, the numerical calculations show that the contribution of the spin flip is four orders of magnitude smaller than that of the dominant radiation.
We also discuss the circular polarization for the generated photon vortices.
\end{abstract}

\maketitle


\section{Introduction}

Allen et al. \cite{Allen92} have proposed optical vortices with a vortex wave, and so far the optical vortices have been studied in fundamental science \cite{Jentschura11a,Hemsing13,Petrillo16,Sherwin17, Peshkov18, Maruyama19a} 
and various applications \cite{Shen19}. 
In addition, Allen et al. \cite{Allen92} have also proposed photon vortices in quantum level.
The wave functions based on Laguerre-Gaussian wave \cite{Maruyama19a} and Bessel wave \cite{Jaurgui05,GamGene,GamGenEx} are known as those for photon vortices.
The generation of the photon vortex was experimentally examined using quantum entanglement \cite{Mair01, Leach02}.
Optical (photon) vortices are considered to be naturally generated in astrophysical sites such as rotating black holes \cite{TTMA11} and strongly magnetized neutrons stars \cite{GamGene}.
Recently, the optical vortex from the M87 black hole was measured \cite{Tamburini20}.

One of the remarkable features of the photon vortex is that it is the eigenstate of the $z$ component of the total angular momentum (zTAM) when the photon propagates along the $z$ axis.
This plays important roles in interactions with quantum objects such as molecules \cite{Babiker02, ACD06} and atoms \cite{Picon10, Afanasev13}, because quantum effects can arise from significant angular momentum transfer.
It has been experimentally demonstrated that the interaction between a photon vortex and a material such as an atom is different from that with a photon described by a plane wave \cite{Schmiegelow16,Lange22}.
%
Furthermore, it is theoretically predicted that one can observe the difference between high-energy photon vortices and plane wave photons in interactions with nuclei \cite{Taira17,Afanasev18,Lu23,Xu24,Maruyama24} and nucleon \cite{Afanasev21} as well as in quantum entanglement \cite{Karlovets23}.
Afanasev et al.~\cite{Afanasev18} have calculated photodisintegration reactions on a deuteron with a photon vortex.
The recent study \cite{Lu23} shows that when a target nucleus exists on a photon vortex propagation axis, excitation of low multipole giant resonances is forbidden from selection rule of angular momentum.
Furthermore, it was presented that the cross sections of photon-induced reactions with photon vortex with zTAM of 2$\hbar$ at specific angles are different from those with zTAM of 1$\hbar$ \cite{Xu24}.

One of the possible methods for generation of high-energy photon vortices at MeV and GeV energies is Compton scattering with high-energy electrons on optical vortices provided from laser and optical devices \cite{Jentschura11a,Stock15}.
When the photon density increases, nonlinear Compton scattering (NCS) may occur, where an electron absorbs several photons and emits a single photon. 
The NCS has been studied for investigation of nonlinear effects of quantum electrodynamics (QED) and it has been experimentally demonstrated using a high-energy electron accelerator and laser \cite{Bula96,Bamber99}.
The recent progress of laser science has enabled demonstration of NCS with multiphoton absorption using electrons accelerated by laser plasma interactions and laser \cite{Yan17}. 
Thus, various calculations including Ferry picture and locally constant crossed field approximation have been developed \cite{Seipt11,Mackenroth11,Krajewska12,Piazza18,Blackburn18,Ilderton19,King21,Gelfer22,Podszus22,Khalaf23}. 
The NCS is also considered to occur in high-intense photon fields in astrophysical sites and may have important roles for astrophysical phenomena such as x-ray pulsars \cite{Arons72}.

Taira et al.~\cite{Taira17} have proposed generation of $\gamma$-ray vortex using NCS with circularly polarized high-intense laser in head-on collision using classical electrodynamics calculation \cite{Taira17}.
This can be understood by a picture that when an electron absorb $N$ photons with the same spin of $+1{\hbar}$, the spins of the $N$ photons are transferred to the total angular momentum $N{\hbar}$ of the emitted photon.
The same scheme in an arbitrary angle between the incident photons and electron was also calculated \cite{Liu20}.
Furthermore, generation of photon vertices using this method has been studied using three-dimensional particle-in-cell simulations developed for high-intense laser-plasma interactions \cite{Chen18, Zhu18, Wang20}. 
Photon vortex generation has been also calculated using Furry picture \cite{Ababekri24}.
These calculations show that the emitted photons have large angular momenta.
It is predicted that the total angular momenta of the generated photons increase up to 10$^{18}$$\hbar$ using laser with an intensity of 10$^{22}$~W/cm$^2$ \cite{Wang20}.

One of the other approaches is calculation using Feynman diagram.
Because it is possible to calculate each step, at which an electron absorbs or emits a photon, in intermediate states  in Feynman diagram with taking into account conservation of angular momentum, one can calculate the energy spectrum of the emitted photons at a fixed photon absorption number.
This is helpful for the understanding of the elementary process of the photon vortex generation.
In the present study, we show photon vortex generation in NCS with circularly polarized photons in the cylindrical coordinate using Feynman method. 
We calculate numerically the energy spectra of photon vortices as a function of the photon absorption number $N$ for various photon densities, corresponding to laser power of (0.5$\--$2)~$\times$~10$^{19}$~W/cm$^2$.

%
%

\section{Theoretical Calculation}

We use the natural unit in the present calculation. 
The absorption probability of a certain number of photons by an electron depends on the photon density in the system.
In the present calculation, the photon density is given as a tunable parameter without  detailed laser pulse structure.
We define the photon density $n_\gamma = N_\gamma / \Omega_R$ for a total photon number of $N_{\gamma}$ in a system of a cylinder with a volume of $\Omega_R = \Omega_T R_z = \pi R_T^2 R_z$, where $R_z$ and $R_T$ are the height and radius of the cylinder, respectively. 
We also note that incident photons have the same value for the $z$ component of the spin angular momentum (zSAM) of $s_{\gamma} = +1$ or $-1$.
%
In the present calculation we assume that an initial electron being the plane wave absorbs circularly polarized $N$ photons being also the plane wave in head-on collision, and emits a photon with a wave function described by Bessel wave as explained later.
There is a problem that when the electron is described by plane wave, it cannot conserve zTAM higher than the electron spin of 1/2 in intermediate states.
To resolve this problem, we introduce the wave function obtained in a cylindrical coordinate for the electron, which is the wave function based upon Bessel wave being the eigenstate of zTAM \cite{Bliokh11, Bliokh17}.

\begin{figure}[htb]
\begin{center}
{\includegraphics[scale=0.6,angle=0]{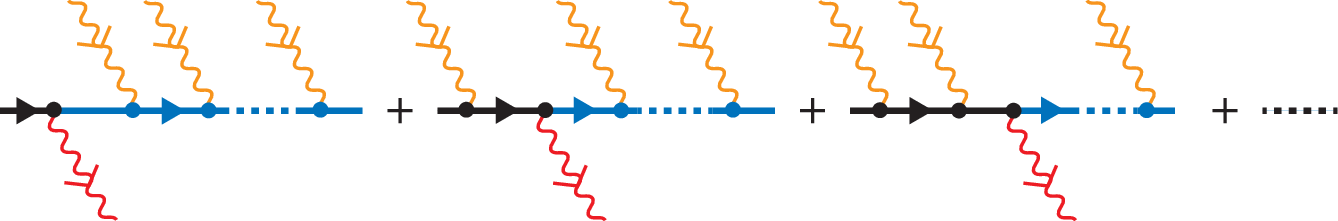}}
\caption{\small
Feynman diagram for nonlinear Compton scattering adopted in the current work. 
The black straight lines stand for the initial electrons described by the plane wave. 
The wave functions of the electrons are changed by Bessel waves (blue color) in Eq.(5) after emitting the vortex photons in Eqs. (2) and (3).
The orange wiggly lines are the incident photons represented by the plane wave, and the red wiggly lines are the photon vortices.
  The first term comes from the time ordering of the photon interaction in the Feynman method.
}
\label{Feynman}
\end{center}
\end{figure}

Figure~\ref{Feynman} 
shows Feynman diagram for the present study.
The vertices indicated by the circles show photon emission or absorption.
The black straight lines stand for the initial electrons described by the plane wave. 
The wave functions of the electrons are changed by Bessel waves (blue color) after emitting the photon vortices.
The orange wiggly lines are the incident photons with circular polarization with the same $s_{\gamma}$, and the red wiggly lines are the photon vortices based on Bessel wave.
In the Feynman diagram (Fig.~\ref{Feynman}),  the first term in the leftmost side indicates the process that the electron emits a photon in the first step and absorbs $N$ photons in the following other steps. 
The second term indicates the process that the electron absorbs a photon in the first step and emits a photon in the second step; the other ($N$$-$1) photons are absorbed in the following steps. 
The third term shows that the electron emits a photon after the absorption of two photons.
If we sum all of the terms, then the electron becomes the Volkov electron in Furry picture.
After the absorption of a photon in any cases, zTAM of the electron increases by 1 originating from the spin of the photon when the incident photons have $s_{\gamma} = +1$.
However, plane-wave electrons cannot conserve zTAM lager than 1/2, because the plane wave is not the eigenstate of the $z$ component of the orbital angular momentum (zOAM).
Thus, we describe the electron after the  emission of a photon by Bessel wave obtained for a free electron in the cylindrical coordinate as explained later to conserve zTAM increased through absorption of multiple photons having the same $s_{\gamma}$.

\noindent
\subsection{wave functions of radiated photons}

%
For photons in the cylindrical coordinate the wave functions based on Bessel wave  has been obtained as a solution of the Klein-Gordon equation \cite{GamGene}.
These wave functions are the eigenstates of the transverse magnetic (TM) state 
and the transverse electric (TE) state at a fixed zTAM of $m$ when the photon propagates along the $z$ axis \cite{Jaurgui05,GamGene,GamGenEx}.
They are written as

\noindent
\begin{minipage}{0.98\hsize}
\begin{equation}
 \tilde{\bA}_m^{(\alpha)} (\vbr,z,e_q) = \vA_m^{(\alpha)} 
 \frac{e^{i (q_z z - e_q t) } }{ \sqrt{\Omega_R} }
\end{equation}
\end{minipage}
with $\alpha$ being TM or TE, and

\noindent
\begin{minipage}{0.98\hsize}
\begin{eqnarray}
\vA_{m}^{(TM)}  (\vbr) &=&  
\frac{1}{2 e_q } \left[ i q_z \left( \tldJ_{m+1} - \tldJ_{m-1}  \right) , 
q_z \left( \tldJ_{m+1} + \tldJ_{m-1} \right),  2 q_T \tldJ_{m}  \right] ,
\nonumber
\label{PhTM}
\\
\vA_{m}^{(TE)}  (\vbr) &=& 
\frac{1}{2 } \left[ i \left( \tldJ_{m+1}  + \tldJ_{m-1}  \right) , 
 \left( \tldJ_{m+1} - \tldJ_{m-1} \right), 0  \right] ,
\label{PhTE}
\end{eqnarray}
\end{minipage}
\begin{equation}
\tldJ_M (q_T \vbr)  = J_M( q_T r) e^{i M \phi} ,
\label{bessel}
\end{equation}
where $J_{M}$ is the $M$-th Bessel function and one of $J_m$, $J_{m-1}$, and $J_{m+1}$ is used as $M$, $e_q$ is the photon energy, and $q_z$ and $q_T$ are the $z$ and transverse components of the momentum with $q_T = \sqrt{e_q^2 - q_z^2}$.
For the discussion we will have later, we define the helicity states as
$\vA_{h = \pm 1} = \left[ \vA^{(TM)} \mp \vA^{(TE)} \right]/\sqrt{2}$.
\\

\noindent
\subsection{wave functions of electrons after photon emission and electron propagator}

We write the electron wave function in the cylindrical coordinate as
${\tilde \psi}(\zr,z) = \psi (\zr) \exp(ip_z z)/\sqrt{R_z}$.
This $\psi$
is a solution of Dirac equation:
\begin{eqnarray}
&& \left\{  {\valpha} \cdot (-i {\bm \nabla}_r ) + \alpha_z p_z + \beta m_e - E \right\}  \psi (\zr) = 0 ,
\end{eqnarray}
where $\valpha$ and $\beta$ are the Dirac matrices, $E$ is the electron energy, and $m_e$ is the electron mass.
The wave functions based on Bessel wave have been known as a solution \cite{Bliokh11, Bliokh17}.
In this study, we use the following wave functions.
The wave functions of the positive and negative energy states are written as
\begin{eqnarray}
\psi_{L,s}^{(+)} (\vbr)  &=& \
\sqrt{\frac{E + m_e}{2 \Omega_T E} } 
\left[ \begin{array}{c} \tldJ_L \chi_s \\ 
\frac{s}{E + m_e} \left( p_z \tldJ_L \chi_s - i p_T \tldJ_{L+s} \chi_{-s} \right) 
 \end{array} \right] ,
\nonumber
\label{PelWf}
\\
\psi^{(-)}_{L,s}  (\vbr) &=& \sqrt{\frac{E + m_e}{2 \Omega_T E} } 
\left[ \begin{array}{c}  
- \frac{s}{E + m_e} \left( p_z \tldJ_L \chi_s - i  p_T \tldJ_{L+s} \chi_{-s} \right) 
\\ \tldJ_L  \chi_s 
 \end{array} \right] ,
\label{NelWf}
\end{eqnarray}
where $\vbr = (\vbr_T, z) = (x,y,z)$, $L$ and $s/2$ are zOAM and zSAM, respectively, $J_L$ is the $L$-th Bessel function as shown in Eq.~(\ref{bessel}), and $p_z$ and $p_T$ are the $z$ and transverse components of the fermion momentum, and $E$ is the energy $E= \sqrt{p_T^2 + p_z^2 + m_e^2}$.
This wave function is neither the eigenstate of zOAM nor zSAM, but it is also the eigenstate of a zTAM of $J$ as expected, where $J$ satisfies the relationship of $J = L + s/2$ ($s=\pm 1$).

The probability of Compton scattering can be calculated from the electron propagator in addition to the wave functions of the electron and photon.
We can obtain the electron propagators from the wave function of the electron in Eq.~(\ref{NelWf}) as follows,

\noindent
\begin{minipage}{0.98\hsize}
\begin{eqnarray}
S_e( \vbr, \vbr^\prime; p_z,  E )
&=&   \Omega_T  \sum_{n} \int \frac{d p_T p_T}{(2 \pi)} 
\left\{  \frac{\psi_{n}^{(+)} (\vbr) \psibar_{n}^{(+)} (\vbr)}{E - E_n + i \delta}
+  \frac{\psi_{n}^{(-)} (\vbr) \psibar_{n}^{(-)} (\vbr)}{E - E_n - i \delta}
\right\}
\label{E-Prop1}
%
%
\end{eqnarray}
\end{minipage}
where $n$ indicates $L$ and $s$, and the energy of the state $n$ is 
$E_n = \sqrt{p_z^2 + p_T^2 + m_e^2}$. 
\\

\noindent
\subsection{Photon production probability}

As the next step, we calculate the production probability of a photon in NCS with circularly polarized photons using the electron propagator.
Although spin polarized beams have been sometimes used in nuclear and particle physics experiments, we consider that there is no spin polarization for the initial electron.
We assume both spin states are included in equal amounts, and we take their average after calculation 
for the initial spin state.
We assume that the initial electron is the plane wave with four momentum of $(E_i ; 0,0, p_{iz})$ and zTAM of $s_i/2$,  
and absorbs $N$ plane-wave photons with momentum $(k ; 0, 0, -k)$ and zTAM of $s_{\gamma}$.
The electron changes to the final state after the emission and absorption.
The electron emits a photon vortex with a zTAM of $K$ having $z$ and transverse components of a momentum of $q_z$ and $q_T$, respectively.
The final electron is the eigenstate of zTAM of $J_f$.
Note that the total number of the steps, at which the electron absorbs or emits a photon, is $(N+1)$ when the electron absorbs $N$ photons and emits one photon.

We explain the more detailed calculation for obtaining the production probability in the NCS with initial photons with the same $s_{\gamma}$.
The production probability in the NCS is written as
\begin{eqnarray}
d P &=& \frac{1}{N!}  (2 \pi)^2 \delta(E_f + e_q - E_i - N k) \delta(p_{fz} + q_z - p_{iz} + N k)
\frac{p_{fT} d p_{fT} d p_{z}}{(2 \pi)^2} \frac{q_T d q_T d q_z }{(2 \pi)^2} 
\nonumber \\ && \times 
\left|
\int d^2 \vbr_{N+1} \cdots d^2 \vbr_1  \psi_f^\dagger (\vbr_{N+1}) 
\right.  \nonumber \\ &&  \quad \times
\left\{ \left( e \sqrt{\frac{N_\gamma}{2 k \Omega_R} }  \Asla_i (\vbr_{N+1}  )\right)
S_e(\vbr_{N+1}, \vbr_{N}) 
 \cdots 
S_e(\vbr_2, \vbr_1)  \left( \frac{e }{\sqrt{2 e_q } }  \Asla_f^* (\vbr_1) \right)
%
\right. \nonumber \\ &&  \qquad 
+ \left( e \sqrt{\frac{N_\gamma}{2 k \Omega_R} }  \Asla_i (\vbr_{N+1}  ) \right)
S_e(\vbr_{N+1}, \vbr_{N})  \cdots
\nonumber \\ && \qquad\qquad\qquad
 \cdots S_e(\vbr_3, \vbr_2)  \left( \frac{e }{\sqrt{2 e_q } }  \Asla_f^* (\vbr_2) \right) 
 S_e(\vbr_2, \vbr_1)  \left( e \sqrt{\frac{N_\gamma}{2 k \Omega_R} } \Asla_i (\vbr_{1})  \right) 
  ~~
 \nonumber \\ && \left.\left. \qquad  
 + \dots \quad \right\}\psi_i (\vbr_1) \right|^2 
\nonumber \\ &=& \frac{e^{2(N+1)} }{2 N! ~e_q} \left( \frac{N_\gamma}{2 k \Omega_R} \right)^N  | W_{if} |^2 (2 \pi)^2 \delta(E_f + e_q - E_i - N k)
\nonumber \\ &&
\qquad \times~
\delta(p_{fz} + q_z - p_{iz} + N k)
   \frac{p_{fT} d p_{fT} d p_{z}}{(2 \pi)^2} \frac{q_T d q_T d q_z }{(2 \pi)^2} ,
\label{PrPrb}
 \end{eqnarray}
\begin{eqnarray}
W_{if} &=&  \sum_{M_{N+1}} \cdots \sum_{M_1}  \left\{ \frac{\cW_A (f, M_{N+1}) \cW_A (M_{N+1}, M_{N} ) \cdots \cW_E (1,i)}{\left(E_f - k -e_q - d_N E_N \right) \cdots (E_i - e_q - d_1 E_1)}
\right. \nonumber \\ && \left. \qquad\qquad
+ \frac{\cW_A (f, M_{N+1}) \cW_A (M_{N+1}, M_{N} ) \cdots \cW_E (2,1) \cW_A(1,i) }{ \left(E_f - k -e_q - d_{N+1} E_{N+1} \right) \cdots (E_i +k - e_q - d_2 E_2) (E_i +k  - d_1 E_1)}
\right. \nonumber \\ && \left. \qquad\qquad + \cdots
 \right\},
\label{TDenom}
\end{eqnarray}
where $A_i$ is the plane wave for the initial photon, the $A_f$ is the Bessel wave for the final photon, $M_n$ is the intermediate state in the $n$-th step, $d_n = +1$ $(-1)$ indicates the positive (negative) energy states in the $n$-th step,  $i$ and $f$ indicate the initial and final states, 
and $\cW_A$ and $\cW_E$ are the matrix elements of the photon absorption and emission,
which are written as
\begin{eqnarray}
\cW_E (k, n) &=& \int d^2 \vbr \psibar_k (\vbr) \gamma_\mu \psi_n (\vbr) A_f^{* \mu} (\vbr) ,
\label{WE}
\\
\cW_A (k, n) &=&  \int d^2 \vbr \psibar_k (\vbr) \gamma_\mu \psi_n (\vbr) A_i^{\mu} (\vbr) .
\label{WA}
\end{eqnarray}

Next we calculate the integration in Eqs.~(\ref{WE},~\ref{WA}).
We define the particle 1 with the plane wave with momentum $p_1 \equiv (E_1, 0, 0, p_{1z})$ as the initial particle
and the particle 2 with the Bessel wave with the transverse momentum $p_{2T}$.
We write the absorption matrix element for the plane wave electron from the initial state to the first step as
\begin{equation}
\cW_A (1,i) = \frac{2 \pi }{ p_{1T} \Omega_T} \delta \left( p_{1T}  \right)
\sqrt{\frac{ (E_i + m_e) (E_1 + m_e) }{ 2 E_i E_1 } }  \cA_p(1,i).
 \end{equation}
$\cA_p$ is written as
 \begin{eqnarray}
 \cA_p (s_2, d_2; s_1,d_1) &=& 
\delta_{s_1, - \sgm} \delta_{s_2, \sgm} \left\{
 \delta_{d_1, d_2}  d_1  s_\gamma
\left[ \frac{p_{1z} }{E_1 + m_e} - \frac{p_{2z} }{E_2 + m_e}  \right]  %
\right.
\nonumber \\ 
&& \qquad\qquad ~ \left. +
 \delta_{d_1, -d_2} \left[ 1 + \frac{p_{1z} p_{2z} }{( E_1 + m_e )( E_2 + m_e )}  \right]  
\right\} .
\label{AbsPl}
\end{eqnarray}
The absorption matrix element for the  Bessel-wave electron in an intermediate state, for example from the $n=1$ step to the $n=2$ step, is written as
\begin{equation}
\cW_A (2,1) = \frac{ 2 \pi }{p_{1T} \Omega_T } \delta \left( p_{1T}  - p_{2T} \right) 
\sqrt{\frac{ (E_1 + m_e) (E_2 + m_e ) }{ 2  E_1 E_2 } } \cA_B (2,1 ).
\end{equation}
$\cA_B$ is written as
%
%
\begin{eqnarray}
&& 
\cA_B( s_2, d_2, L_2; s_1, d_1, L_1 )
\nonumber \\ 
&=& 
\delta_{d_1,d_2} d_1 s_\gamma \left\{
\delta_{s_1, - \sgm} \delta_{s_2, \sgm}  \delta_{L_2, L_1}  
\left( - \frac{p_{1z}}{E_1 + m_e} + \frac{p_{2z}}{E_2 + m_e}  \right)
 \right. \nonumber \\ && \left. \qquad \qquad
- i  \delta_{s_1, \sgm} \delta_{s_2, \sgm}  \delta_{L_2, L_1 + \sgm}  \frac{p_{1T} }{E_1 + m_e} 
- i  \delta_{s_1, - \sgm} \delta_{s_2, - \sgm}  \delta_{L_2, L_1 + \sgm} \frac{p_{2T}}{E_2 + m_e}     
\right\}
\nonumber \\  && 
+ \delta_{d_1, -d_2} \left\{ \delta_{s_1, - \sgm} \delta_{s_2, \sgm} \delta_{L_2, L_1}  \left[ 1 + \frac{ p_{1z} p_{2z}}{(E_1 + m_e) (E_2 + m_e)} \right] 
\right. \nonumber \\ && \qquad\qquad
 - \delta_{ s_1,  \sgm}  \delta_{ s_2,  - \sgm}  \delta_{L_2, L_1 - 2\sgm} \frac{p_{1T} p_{2T} }{(E_1 + m_e) (E_2 + m_e)}  
 \nonumber \\ && \qquad\qquad
+ \delta_{L_2, L_1 + \sgm }  \left[  \delta_{s_1, \sgm} \delta_{s_2, \sgm} \frac{i p_{1T} p_{2z} }{(E_1 + m_e) (E_2 + m_e)}
\right.  \nonumber \\ && \left.\left.  \qquad \qquad\qquad\qquad\quad
-  \delta_{s_1, -\sgm} \delta_{s_2, -\sgm}  \frac{i p_{1T}  p_{1z} }{(E_1 + m_e) (E_2 + m_e)}  \right] 
\right\} ,
\label{EmiPI}
\end{eqnarray}
Next, we consider the photon emission. The detailed expression of the matrix element for the photon emission, for example from the $n=1$ step to the $n=2$ step, is written as
\begin{equation}
\cW_E = \frac{(2 \pi)}{\Omega_T} (-)^{K}
\sqrt{\frac{(E_1 + m)(E_2 + m)}{ 2 E_1 E_2 } } \frac{\delta(p_{2T} - q_T) }{q_T}
\cM_E 
\end{equation} 
with $E_2 = \sqrt{ (p_{1z}^2 - q_z)^2 + p_{2T}^2 +m_e^2}$ and $\cM_E$ have been written in Eq.~(\ref{MTM-TE}).
\begin{eqnarray}
\cM_E^{(TM)}
 = \frac{1}{ \sqrt{2} e_q} \left[q_z \left( {\cE}_{+1} - {\cE}_{-1} \right) + q_T {\cE}_0 \right] , \quad
&&
\cM_E^{(TE)} = \frac{1}{ \sqrt{2} } \left[ {\cE}_{+1} + {\cE}_{-1} \right] ,
\label{MTM-TE}
\end{eqnarray}
where
\begin{eqnarray}
&&\cE_{s =\pm1} (h_2, d_2, L_2; h_1,d_1)  
\nonumber  \\ &=& ~
 \delta_{d_1, d_2} d_1 \left\{
i \left( \frac{ p_{1z} }{E_1 + m} - \frac{ p_{2z} }{E_2 + m} \right)  \delta_{h_1, s} \delta_{h_2, -s} \delta_{ L_2, - K +s } 
- \frac{ q_{T}  }{E_2 + m} \delta_{s_1, s}  \delta_{s_2, s}  \delta_{ L_2, - K}    
\right\}
%
\nonumber \\ &&  
+ \delta_{d_1, -d_2}  \left\{ 
i \left[ 1 + \frac{p_{1z} p_{2z} }{(E_1 + m)(E_2 + m)} \right] \delta_{s_1, s} \delta_{h_2, -s} \delta_{L_2, -K+s} 
\right. \nonumber \\ && \left. \qquad\qquad
+ \frac{ q_{T} p_{1z} }{(E_1 + m)(E_2 + m)} \delta_{s_1, s} \delta_{s_2, s} \delta_{L_2, -K}  
\right\} ,
\label{EmiCa1}
\end{eqnarray}
\begin{eqnarray}
 && \cE_0 (h_2, d_2, L_2; h_1,d_1) 
 %
\nonumber  \\ &=& 
\delta_{d_1, d_2} d_1 
\left\{ \left( \frac{ p_{1z} }{E_1 + m}  + \frac{ p_{2z} }{E_2 + m}  \right) \delta_{s_1, s_2} \delta_{L_2, -K}  
 - \frac{ i q_{T} }{E_2 + m} \delta_{s_1, -s_2 }  \delta_{L_2, -K + s_1}  
\right\}  \qquad\quad
%
\nonumber  \\ && 
+ \delta_{d_1, -d_2} 
h_1  \left\{ \left[ 1 - \frac{ p_{1 z}  p_{2 z}  }{(E_1 + m)(E_2 + m)} \right] \delta_{s_2, s_1} \delta_{ L_2, - K}  
\right. \nonumber  \\ && \left. \qquad\qquad\qquad\qquad
+ \frac{ i q_{T}  p_{1 z} }{(E_1 + m)(E_2 + m)} \delta_{ s_2, -s_1}  \delta_{ L_2, - K - s_2} 
\right\} .
\label{EmiCa0}
\end{eqnarray}

The summation for the $n$-th intermediate states are given by
\begin{equation}
\sum_{M_n} \equiv \Omega_T (2 \pi) \int d p_{nT} p_{nT}
\sum_{d_n = \pm 1}\sum_{s_n=\pm 1} \sum_{L_n} .
\end{equation}
If the intermediate state is the plane wave along the $z$ direction, the summation becomes
\begin{equation}
\sum_{M_n} \equiv \sum_{d_n = \pm 1}\sum_{s_n = \pm 1} \sum_{L_n} .
\end{equation}
Therefore  the $W_{if}$ in Eq.~(\ref{TDenom}) is written as
\begin{equation}
W_{if} = W_{if}^{(1)} +  W_{if}^{(2)}
\end{equation}
with
\begin{eqnarray}
W_{if}^{(1)} &=& \sum_{M_1} \cdots \sum_{M_{N+1}} \cW_A(f, N+1) \cdots \cW_A(2,1) \cW_E(1,i) 
\nonumber \\ 
&=&  \frac{(2 \pi)}{ \Omega^{(N+3)/2} q_T } \sqrt{ \frac{(E_f+m_e)(E_i+m_e)}{4 E_i E_f} } \cM_1 ,
%
%
%
\end{eqnarray}
%
\begin{eqnarray}
W_{if}^{(2)} &=& \sum_{M_1} \cdots \sum_{M_{N+1}} \cW_A(f, N+1) \cdots \cW_E(2,1) \cW_A(1,i) 
\nonumber \\ 
&=&  \frac{(2 \pi)}{ \Omega^{(N+3)/2} q_T } \sqrt{ \frac{(E_f+m_e)(E_i+m_e)}{4 E_i E_f} } \cM_2 ,
%
%
%
\end{eqnarray}
where $\cM_{1}$ and $\cM_{2}$ are

\noindent
\begin{minipage}{0.98\hsize}
\begin{eqnarray}
\cM_1 &=& \left[ \prod_{n=1}^{N} \left( \frac{E_n + m_e}{2 E_n} \right) \right] 
\sum_{s_1 \cdots s_{N}} \sum_{d_1 \cdots d_{N}}
\frac{ \cA_B(f,N) \cdots \cA_B (2,1) \cM_E(1,i) }{\left(E_f - k - d_{N} E_{N} \right) \cdots (E_i - e_q - d_1 E_1)} , 
\label{M-El1}
\\
\cM_2 &=& \left[ \prod_{n=1}^{N} \left( \frac{E_n + m_e}{2 E_n} \right) \right] 
\sum_{s_1 \cdots s_{N} } \sum_{d_1 \cdots d_{N}}
\frac{ \cA_B(f,N) \cdots \cM_E(2,1)  \cA_p (1,i) }{\left(E_f - k - d_{N} E_{N} \right) \cdots (E_i + k - d_1 E_1)} .
\label{M-El2}
\end{eqnarray}
\end{minipage}
%
In the above equations, the energy of the $n$-th step $E_n$ is independent of $s_n$ and $d_n$
in $\cM_1$ and  $\cM_2$, respectively. 
\\

%
%
We finally obtain the production probability using the above equations, which is written as

\noindent
\begin{minipage}{0.98\hsize}
\begin{eqnarray}
d P  &=&
\frac{e^{2(N+1)} }{4 \pi N! (2 k)^N} n_\gamma^N \frac{(E_f+m_e)(E_i+m_e)}{4 e_q E_i E_f} 
 \nonumber \\ && \quad \times 
\sum_{s_i, s_f} |\cM_1 + \cM_2|^2
\delta (E_f + e_q - E_i - N k)  1q_T d q_T d q_z .
\label{PrPrb2}
\end{eqnarray}
\end{minipage}
%
By performing the integration, we can rewrite the above equation as

\noindent
\begin{minipage}{0.98\hsize}
\begin{eqnarray}
 \frac{d P}{d e_q }  &=&  
\frac{e^{2(N+1)} }{ 16 \pi N! }  \left( \frac{n_\gamma}{2 k} \right)^N
\frac{(E_f+m_e)(E_i+m_e)}{ E_i ( p_{iz} - Nk)  } \sum_{s_i, s_f}
|\cM_1 + \cM_2|^2 .
%
\label{PrdPr}
\end{eqnarray}
\end{minipage}

The first term in Fig.~\ref{Feynman} shows that an electron emits a photon in the first step; it changes to the Bessel-wave electron, which can absorb multiple photons in the following intermediate steps.
The second term shows that an electron with $s_i/2 = -1/2$ absorbs a photon of $s_{\gamma} = +1$ through a spin-flip process  $-1/2 \rightarrow +1/2$ in the first step.
However, it cannot absorb further photons in the following steps.
Thus, in the second step the electron should emit a photon and change to the Bessel-wave electron to absorb further photons.
Therefore, the photon emission must be done in the first or the second step and the only first and second terms in Fig.~\ref{Feynman} remain.
Note that if $s_i/2 = +1/2$ the emitted photon could obtain an additional angular momentum from the spin flip of $+1/2 \rightarrow -1/2$ in the first step.
This indicates that when an electron absorbs $N$ photons the radiated photon could have zTAM of $(N+1)$ in addition to $N$.
\\

\subsection{Numerical calculation}

We here demonstrate the energy spectrum calculation using the presently obtained solution under specific conditions.
To observe photon vortices with a fixed zTAM we need to set the main axis, although the initial plane-wave photons and electron do not have such axis.
Thus, we set the main axis by fixing the OAM of the final electron
and we take only the contributions from $L_f = 0$, 
which correspond to photons scattering along the $z$ axis.
In contrast, if we take non-zero zOAM for the final electron, the photon wave function is mixed by different zTAM states.
In the case of $L_f = 0$, the final electron is described by Bessel wave without any additional angular momentum 
so that we could not expect the effect cased by the large intrinsic angular momentum of the electron vortex.

\begin{figure}[htb]
\begin{center}
{\includegraphics[scale=0.55,angle=270]{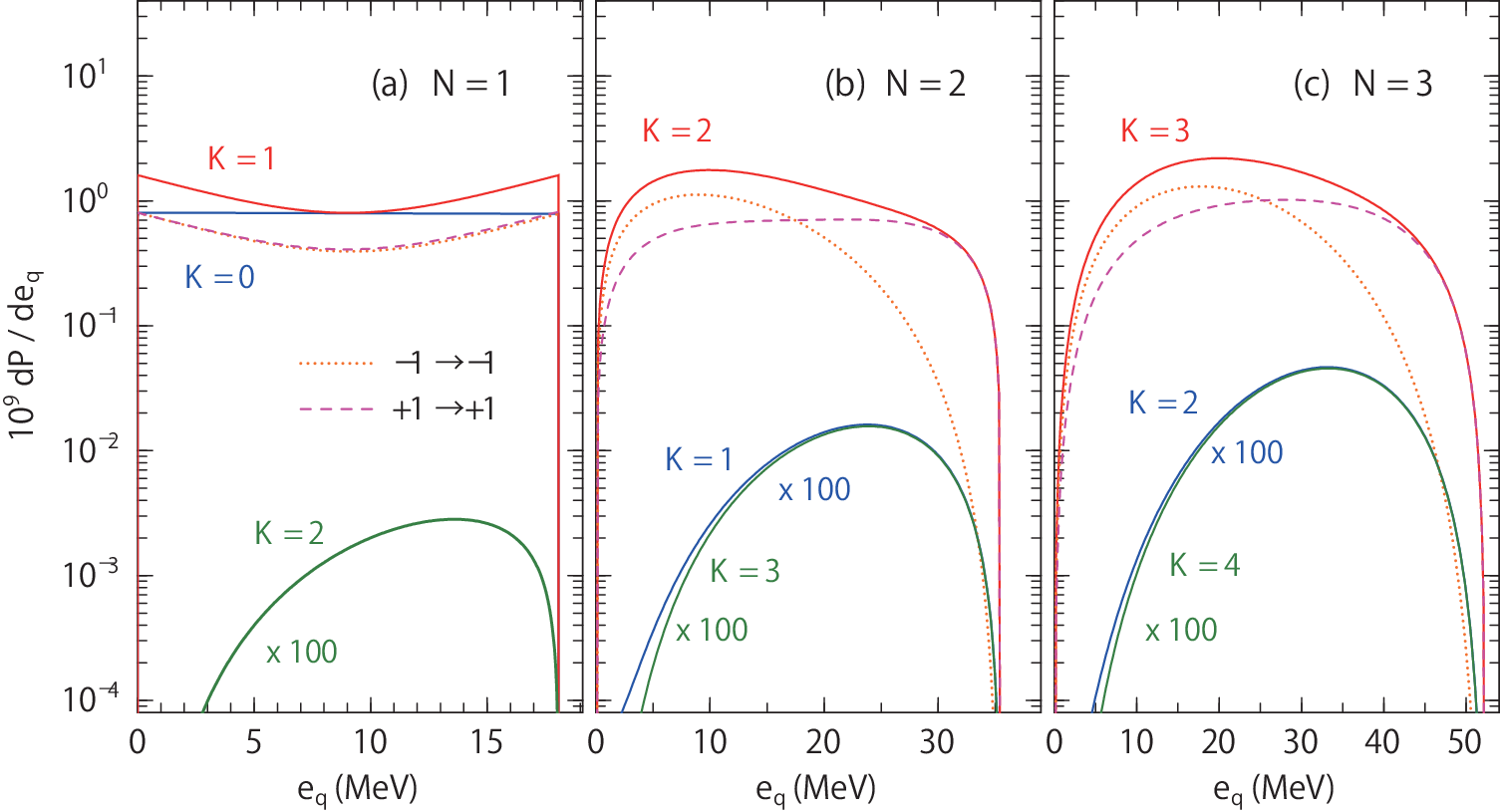}}
\caption{\small
Differential photon production probabilities for one photon absorption ($N=1$) (a), 
two photon absorption  ($N=2$)  (b), and three photon absorption  ($N=3$)  (c)  with $E_i - m_e = 1$~GeV.
The blue, red, and green solid lines represent the results 
when $K = N -1$, $K = N$ and $K = N+1$, respectively.
The red dashed and dotted lines show the contributions from $s_i = s_f = -1$ and
 $s_i = s_f = +1$ with $K = N$, respectively. 
 }
\label{PhAbs}
\end{center}
\end{figure}

We assume the initial photon energy of $k=1.2$~eV and the electron kinetic energy of 1~GeV in the following calculations.
We set $s_{\gamma} = +1$ for the initial photons and the photon density of 1.5${\times}10^{6}$~nm$^{-3}$ corresponding to the laser intensity of 0.9$\times$10$^{19}$~W/cm$^2$.
Figure~\ref{PhAbs} shows the photon production probabilities in the NCS as functions of the emitted photon energy for photon absorption of $N = 1{\--}3$.    
The strength at $K = N$  is dominant in each channel of the absorption photon number, 
and the strength of $K = N {\pm} 1$ for $N = 2 {\--} 3$ is four orders of magnitude smaller than that of $K = N$.
Note that even if an electron absorbs only one photon, photon vortex with $K = 2$ could be produced through electron spin flip although its strength is also four orders of magnitude smaller than that of the fundamental radiation.
This shows that almost all photons generated by NCS with circularly polarized photons are photon vortices with zTAM $\ge$ 2.
The photons with $K = N {\pm} 1$ are generated by spin flip of the initial electron, 
but its probabilities are negligibly small. 
The two modes of $s_i = s_f = -1$ and $s_i = s_f = +1$ are dominant in electron spin channels.

\begin{figure}[hbt]
\begin{center}
{\includegraphics[scale=0.5,angle=0]{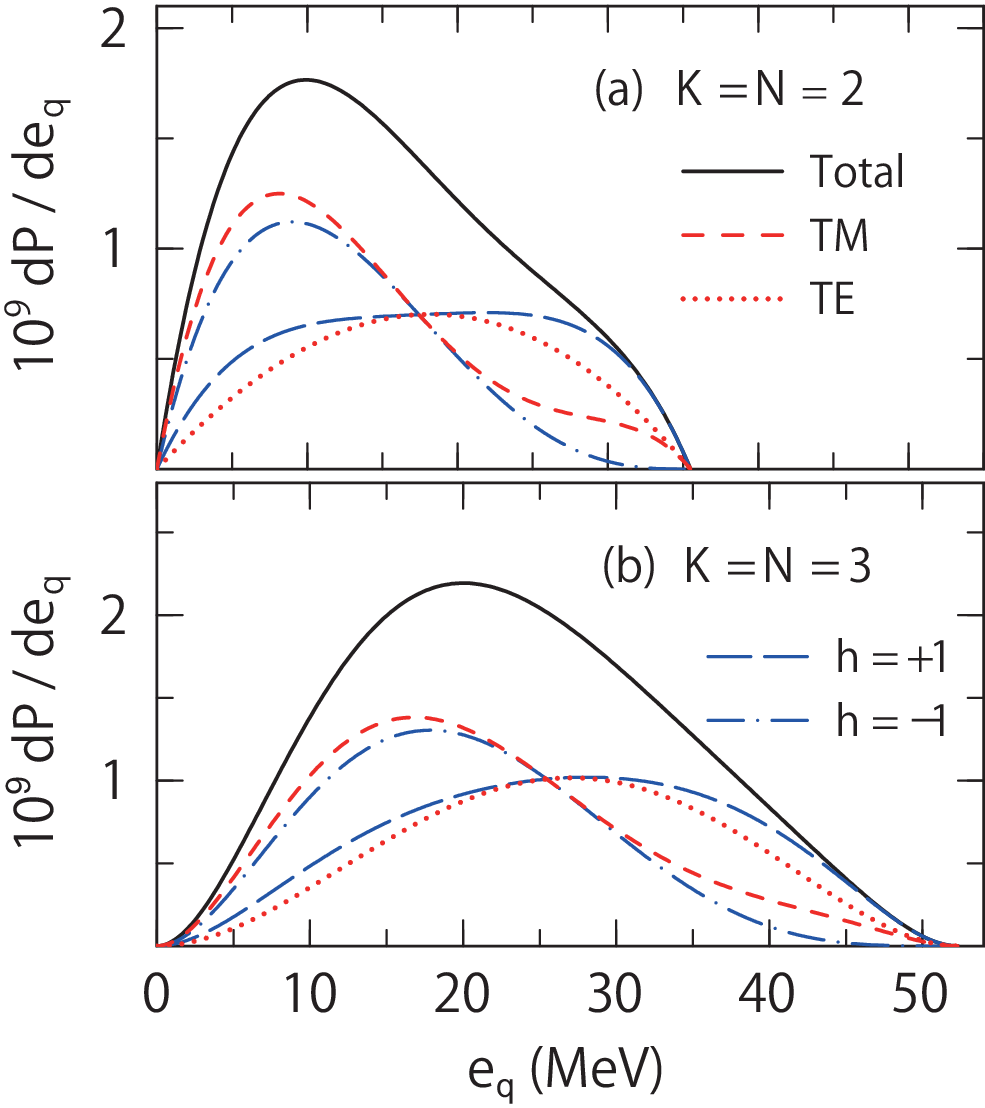}}
\caption{\small
Differential  photon production probabilities of $10^{9} d P/de_q$  when $K = N =2$ (a)
and $K = N =3$ (b) with $E_i - m_e = 1$~GeV.
The red dashed and blue dotted lines represent the contributions 
from the emission channels of  the TM and TE modes, respectively.
The green and red long-dashed lines indicate the contribution
from the helicities $h=+1$ and $h=-1$ states, respectively. 
}
\label{ChCmp}
\end{center}
\end{figure}

As explained previously, the photon wave functions are not the eigenstate of the helicity but that of the TM and TE modes.
Figure~\ref{ChCmp} shows the contribution of $d P/de_q$ from the TM and TE modes when $K = N =2$ and $3$.
The spectra for the two modes are different.
The TM mode is higher than the TE mode in the high energy region, whereas its relationship is inverted in the low energy region.
We also show the fraction of each helicity.
When one of the two helicity states dominates, a strong circular polarization is observed.
At present, it is possible to measure circular (linear) polarization of x rays from the Universe by detectors located outside the atmosphere \cite{Wiersema14}.
Figure~\ref{ChCmp}  shows that near the maximum energy for each $K$ value the $h = +1$ states dominate, but as the energy decreases the fraction of the $h = -1$ states increases.
This suggests that when one can observe the circular polarization close to the maximum energy for a fixed $K$ value.
However, because photon vortices with different $K$ values are mixed at a given energy as discussed later,
only weak circular polarization can be observed over a wide energy range.
This is similar to the situation for photon vortex generation by harmonic radiations in synchrotron radiations from spiral motion of an electron under a uniform magnetic field \cite{GamGene},
where the same Bessel wave in  Eq.~(\ref{PhTE}) is also obtained for the photons.

\begin{figure}[htb]
\begin{center}
{\includegraphics[scale=0.55]{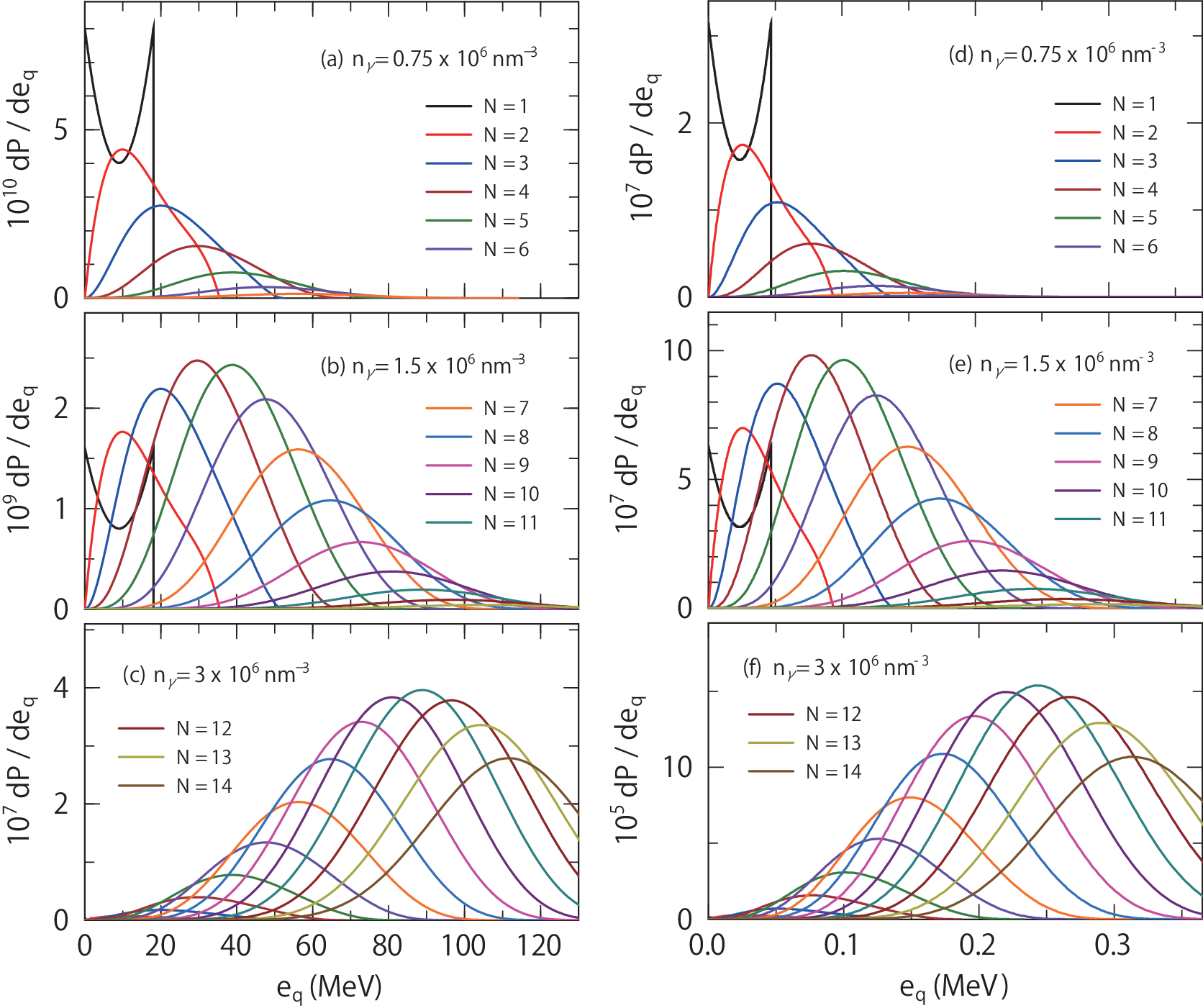}}
\caption{\small
Differential production probabilities for the nonlinear Compton scattering with $N=K$ for the photon density of $n_{\gamma}$ = 0.75$\times$10$^{6}$~$nm^{-3}$ (a),  1.5$\times$10$^{6}$~$nm^{-3}$ (b),  3$\times$10$^{6}$~$nm^{-3}$(c)
when $E_i - m_e = 1$~GeV
and 
of $n_{\gamma}$ = 0.75$\times$10$^{6}$~$nm^{-3}$ (d),  1.5$\times$10$^{6}$~$nm^{-3}$ (e),  3$\times$10$^{6}$~$nm^{-3}$(f)
when $E_i - m_e = 50$~MeV
}
\label{PhEp}
\end{center}
\end{figure}

In the present scheme, we can calculate NCS in an intermediate photon density where NCS does not dominate completely.
Figure~\ref{PhEp} shows the photon production probabilities with photon densities of $n_{\gamma} = 0.75{\times}10^{6}$, 1.5${\times}10^{6}$, and 3${\times}10^{6}$~nm$^{-3}$, corresponding to laser intensities of approximately 0.45$\times$10$^{19}$, 0.9$\times$10$^{19}$, and 1.8$\times$10$^{19}$~W/cm$^{2}$, respectively, for electron kinetic energies of 50~MeV and 1~GeV.
These laser intensities could be provided from many laser facilities at present.
In the lowest photon density corresponding to 0.45$\times$10$^{19}$~W/cm$^2$ (0.75${\times}10^{6}$~nm$^{-3}$), the fundamental radiation by Compton scattering  only with one photon is the most strong, but in the energy region higher than the maximum energy of the fundamental radiation, only photon vortices are generated.
The average $K$ of the radiated photons increases with increasing the photon density.
When the laser intensity is higher than  0.9$\times$10$^{19}$~W/cm$^2$ (1.5${\times}10^{6}$~nm$^{-3}$) almost all generated photons are photon vortices with an average $K$ higher than 3$\hbar$.
In the case of laser intensity of 1.8$\times$10$^{19}$~W/cm$^{2}$  (3${\times}10^{6}$~nm$^{-3}$), the average $K$ becomes as high as 11$\hbar$.
The present result shows that the present method is effective for calculation of NCS with circular polarized photons
in the assumed photon densities, corresponding to the absorption photon number of $N=2{\--}20$.
The absorption photon number increases with increasing the initial photon density.
When the absorption photon number becomes much larger than 100,
other methods such as Furry picture
\cite{Seipt11,Mackenroth11}
and locally constant field approximation
\cite{Piazza18,Ilderton19} are more suitable rather than the present approach.

\begin{figure}[htb]
\begin{center}
{\includegraphics[scale=0.6,angle=270]{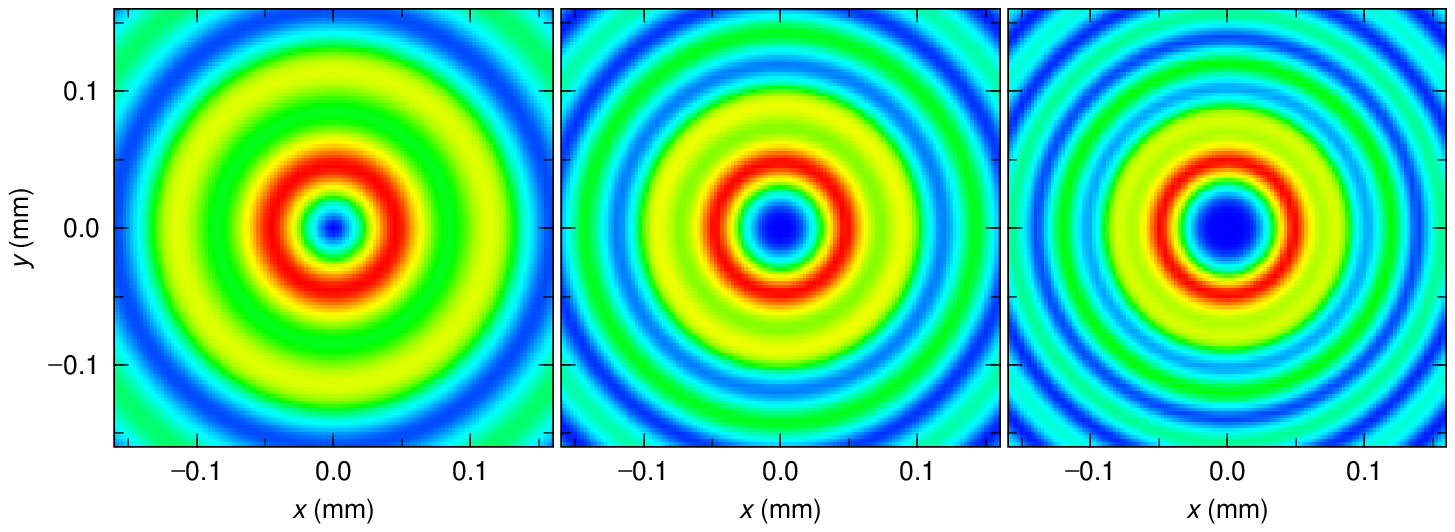}}
\caption{\small
Density distribution of emitted photon per second as a function of the distance from the rotation
axis at 10~MeV (a), 20~MeV (b), and 30~MeV (c) when $E_i - m_e= 1$~GeV.
In these energies, photon vortices with K = 2, 3, and 4 dominate, respectively.
}
\label{Density}
\end{center}
\end{figure}

Figure \ref{Density} shows the number density of the emitted photons as a function of $r_T$ in the $x$-$y$ plane,
and shows multiple rings around the $z$ axis originating from the Bessel wave.
These distributions are written as superposition of the three Bessel function $J_m(r_T) (m = N-1, N, N+1)$ although $J_N$ dominates.
Because the Bessel-wave photons propagate parallel along the $z$ axis, the distributions do not depend on the distance from the scattering point.

\section{Summary}

In the present study, we have calculated photon vortex generation by NCS with high-energy electrons on circularly polarized photons using Feynman diagram.
We have calculated each step in intermediate states taking into account conservation of angular momentum.
The energy spectrum of the emitted photons at a fixed photon absorption number is obtained.
The result is consistent with the previously predicted picture that when a photon absorbs $N$ photons it emits a photon with a zTAM of $N$.
The result shows that photons with a zTAM of ($N-1$) and ($N+1$), which are generated by the spin flip of the initial electron, are also generated, but the numerical calculation suggests that their contributions are four orders of magnitude smaller than that of the dominant photons with a zTAM of $N$.
Even if an electron absorbs only one photon, the electron may emit a photon vortex with a zTAM of 2, albeit with very low probability.
The dominant process corresponds to the process where an electron first emits a photon and subsequently absorbs several photons.
The calculated result also shows that one can observe only weak circular polarization for the radiated photons.

\acknowledgements

This work was supported by Grants-in-Aid for Scientific Research of 
JSPS (JP20K03958, JP22H03881, JP22H01239, JP24K07057).
M.~K.~C. work was supported by the National Research Foundation of Korea (Grant
Nos. NRF-2020R1A2C3006177 and NRF-2021R1A6A1A03043957).
T.~K. is supported in part by the National Key R\&D Program of China
(2022YFA1602401) and the National Natural Science Foundation of China
(No. 12335009 and No. 12435010).

\end{document}